\documentclass[a4paper, 12pt, oneside]{article}
\usepackage[cp1251]{inputenc}
\usepackage[english, russian]{babel}
\usepackage{ graphics,amsfonts, amsmath,bm,amssymb, array, eucal}
\usepackage[dvips]{graphicx}
 {

\usepackage{indentfirst}

\begin{document}
\thispagestyle{empty} \large
\renewcommand{\abstractname}{}
\renewcommand{\abstractname}{Abstract }
\renewcommand{\refname}{\begin{center} REFERENCES\end{center}}
\newcommand{\mc}[1]{\mathcal{#1}}
\newcommand{\E}{\mc{E}}

 \begin{center}
\bf Transverse electric conductivity and dielectric permeability in quantum
non-degenerate and maxwellian collisional plasma with variable
collision frequency in Mermin's approach
\end{center}\medskip
\begin{center}
  \bf A. V. Latyshev\footnote{$avlatyshev@mail.ru$} and
  A. A. Yushkanov\footnote{$yushkanov@inbox.ru$}
\end{center}\medskip

\begin{center}
{\it Faculty of Physics and Mathematics,\\ Moscow State Regional
University, 105005,\\ Moscow, Radio str., 10--A}
\end{center}\medskip

\begin{abstract}
Formulas for transverse conductance and dielectric permeability in quantum
non-degenerate and Maxwellian collisional plasma
with arbitrary variable collision frequency
in Mermin's approach are deduced.
Frequency of collisions of particles depends arbitrarily on a wave vector.
The special case of frequency of collisions
proportional to the module of a wave vector is considered.
The graphic analysis of the real and imaginary parts  of
dielectric function is made.

{\bf Key words:} Klimontovich, Silin, Lindhard, Mermin, quantum
collisional plasma, conductance, non-degenerate plasma, Maxwellian plasma.

PACS numbers: 03.65.-w Quantum mechanics, 05.20.Dd Kinetic theory,
52.25.Dg Plasma kinetic equations.
\end{abstract}

\begin{center}
{\bf 1. Introduction}
\end{center}

In  Klimontovich and Silin's work  \cite{Klim} expression
for longitudinal and trans\-verse dielectric permeability of quantum
collisionless plasmas has been re\-cei\-ved.

Then in Lindhard's work \cite{Lin} expressions
has been received  also for the same characteristics of quantum
collisionless plasma.

By Kliewer and Fuchs \cite{Kliewer} it has been shown, that
direct generalisation of formulas of Lindhard  on a case of collisionless
plasmas, is incorrectly.
This lack for the longitudinal dielectric
permeability has been eliminated in work of Mermin \cite{Mermin} for
collisional plasmas.
In this work of Mermin \cite{Mermin} on the basis of the analysis
of a nonequilibrium matrix
density in $ \tau $-approach expression for
longitudinal dielectric permeability of quantum collisional plasmas
in case of constant frequency of collisions of particles of plasma
has been announced.

For collisional plasmas correct formulas longitudinal and transverse
electric conductivity and dielectric permeability are received
accordingly in works \cite{Long} and \cite{Trans}. In these works
kinetic  Wigner---Vlasov---Boltzmann equation
in relaxation approximation in coordinate space was used.

In work \cite{Trans2} the formula for the transverse electric
conductivity of quan\-tum collisional plasmas with use of the kinetic
Shr\"{o}dinger---Boltzmann equation in Mermin's approach  (in space of
momentum) has been deduced.

In work \cite{Long2} the formula for the longitudinal dielectric
permeability of quantum collisional plasmas with use of the kinetic
Shr\"{o}dinger---Boltzmann equation in approach of Mermin (in space of
momentum) with any variable frequency of collisions depending from
wave vector  has been deduced.

In our work \cite{Lat2007} formulas for longitudinal and transverse
electric con\-duc\-ti\-vity in the classical collisional
gaseous (maxwellian) plasma with frequency of collisions
of plasma particles proportional to the
module particles velocity  have been deduced.

Research of
skin-effect in classical collisional gas plasma with frequency
of collisions proportional to the module particles velocity
has been carried out in work \cite{Lat2006}.

In our works \cite{Long3} and \cite{Long4} dielectric permeability
in quantum collisional plasma with frequency of collisions
proportional to the module
of a wave vector has been investigated. The case of degenerate plasmas
was studied in work \cite{Long3}. The case of non-degenerate
and maxwellian plasmas has been investigated in work \cite{Long4}.

Our work \cite{Trans3} is devoted to transverse
conductivity and permeability in quantum collisional
plasma with  variable frequency of collisions.
In the same work the case вырожденной plasmas is considered.

Let's notice, that interest to research of the phenomena
in quantum plasma grows in last years \cite{Manf} -- \cite{Ropke}.

In the present work formulas for transverse conductivity and dielectric
permeability in quantum
non-degenerate and maxwellian collisional plasma  with arbitrary
vari\-able collision frequency
in Mermin's approach are dedu\-ced.
Frequency of collisions of particles depends arbitrarily on a wave vector.
This work is continuation of our article \cite{Trans3}.
The special case of frequency of collisions
proportional to the module of a wave vector is considered.
The graphic analysis of the real and imaginary parts  of
dielectric function is made.

\begin{center}
\bf 2. Transversal electric conductivity and dielectric permeability
\end{center}

In work \cite{Trans3} we receive the following expression of an invariant form
for the transversal electric con\-duc\-ti\-vity
$$
\sigma_{tr}(\mathbf{q},\omega,\bar\nu)=
\dfrac{ie^2N}{m\omega}\Bigg[1+\dfrac{\hbar^2}{8\pi^3mN}
\int \Xi(\mathbf{k,k-q})(f_{\bf k}-f_{\bf k-q}){\bf k}_\perp^2 d\mathbf{k}
\Bigg].
\eqno{(2.1)}
$$

Here
$$
\Xi({\bf k},{\bf k-q})=\dfrac{\E_{{\bf k}}-\E_{{\bf k-q}}-
i\hbar\bar\nu({\bf k},{\bf k-q})}
{(\E_{{\bf k}}-\E_{{\bf k-q}})\{\E_{{\bf k}}-\E_{{\bf k-q}}-
\hbar [\omega +i\hbar\bar\nu({\bf k},{\bf k-q})]\}},
$$ \medskip
$$
\bar \nu({\bf k},{\bf k-q})=\dfrac{\nu({\bf k})+\nu({\bf k-q})}{2}.
$$

Let's take advantage of definition of transversal dielectric permeability
$$
\varepsilon_{tr}(\mathbf{q},\omega,\nu)=
1+\dfrac{4\pi i}{\omega}\sigma_{tr}(\mathbf{q},\omega,\nu).
\eqno{(2.2)}
$$

Taking into account (2.1) and equality (2.2) we will write expression
for the transversal dielectric permeability
$$
\varepsilon_{tr}(\mathbf{q},\omega,\bar\nu)=1-\dfrac{\omega_p^2}{\omega^2}
\Big[1+\dfrac{\hbar^2}{8\pi^3mN}\int \Xi(\mathbf{k,q})(f_{\mathbf{k}}-
f_{\mathbf{k-q}})\mathbf{k}_\perp^2 d\mathbf{k}\Big].
\eqno{(2.3)}
$$

Here $\omega_p$ is the plasma (Langmuir) frequency,
$\omega_p^2=4\pi e^2N/m$.

Let's notice, that the kernel from subintegral expression from (2.1) can be
we can present in the form of decomposition on partial fractions
$$
\Xi(\mathbf{k,q})\equiv \dfrac{\E_{\mathbf{k}}-\E_{\mathbf{k-q}}-i\hbar \bar\nu({\bf k},{\bf k-q})}
{(\E_{\mathbf{k}}-\E_{\mathbf{k-q}})\{\E_{\mathbf{k}}-\E_{\mathbf{k-q}}-\hbar
[\omega+i\bar\nu({\bf k},{\bf k-q})]\}}=
$$
$$
=\dfrac{i \bar\nu({\bf k},{\bf k-q})}{\omega+i\bar\nu({\bf k},{\bf k-q})}\cdot
\dfrac{1}
{\E_{\mathbf{k}}-\E_{\mathbf{k-q}}}+
$$
$$+
\dfrac{\omega}{\omega+i\bar\nu({\bf k},{\bf k-q})}\cdot\dfrac{1}
{\E_{\mathbf{k}}-\E_{\mathbf{k-q}}-\hbar [\omega+i\bar\nu({\bf k},{\bf k-q})]}.
$$

Hence, for transversal electric conductivity and dielectric
permeability we have explicit representations
$$
\sigma_{tr}(\mathbf{q},\omega,\bar\nu)=
\dfrac{ie^2N}{m\omega}\Bigg[1+\dfrac{\hbar^2}{8\pi^3mN}
\int \dfrac{i \bar\nu({\bf k},{\bf k-q})}{\omega+i\bar\nu({\bf k},{\bf k-q})}
\cdot\dfrac{f_{\bf k}-f_{\bf k-q}}{\E_{{\bf k}}-\E_{{\bf k-q}}}{\bf k}_\perp^2
d{\bf k}+
$$
$$
+\dfrac{\hbar^2\omega}{8\pi^3mN}\int
\dfrac{1}{\omega+i\bar\nu({\bf k},{\bf k-q})}\cdot
\dfrac{(f_{\bf k}-f_{\bf k-q}){\bf k}_\perp^2d{\bf k}}{\E_{{\bf k}}-\E_{{\bf k-q}}
-\hbar [\omega+i\bar\nu({\bf k},{\bf k-q})]}\Bigg],
\eqno{(2.4)}
$$\medskip
and
$$
\varepsilon_{tr}(\mathbf{q},\omega,\bar\nu)=1-\dfrac{\omega_p^2}{\omega^2}
\Bigg[1+\dfrac{\hbar^2}{8\pi^3mN}
\int \dfrac{i \bar\nu({\bf k},{\bf k-q})}{\omega+i\bar\nu({\bf k},{\bf k-q})}
\cdot\dfrac{f_{\bf k}-f_{\bf k-q}}{\E_{{\bf k}}-\E_{{\bf k-q}}}{\bf k}_\perp^2
d{\bf k}+
$$
$$
+\dfrac{\hbar^2\omega}{8\pi^3mN}\int
\dfrac{1}{\omega+i\bar\nu({\bf k},{\bf k-q})}\cdot
\dfrac{(f_{\bf k}-f_{\bf k-q}){\bf k}_\perp^2d{\bf k}}{\E_{{\bf k}}-\E_{{\bf k-q}}
-\hbar [\omega+i\bar\nu({\bf k},{\bf k-q})]}\Bigg].
\eqno{(2.5)}
$$\medskip

If to enter designations
$$
J_{\bar\nu}=\dfrac{\hbar^2}{8\pi^3mN}
\int \dfrac{i \bar\nu({\bf k},{\bf k-q})}{\omega+i\bar\nu({\bf k},{\bf k-q})}
\cdot\dfrac{f_{\bf k}-f_{\bf k-q}}{\E_{{\bf k}}-\E_{{\bf k-q}}}{\bf k}_\perp^2
d{\bf k}
$$
and
$$
J_\omega=\dfrac{\hbar^2\omega}{8\pi^3mN}\int
\dfrac{1}{\omega+i\bar\nu({\bf k},{\bf k-q})}\cdot
\dfrac{(f_{\bf k}-f_{\bf k-q}){\bf k}_\perp^2d{\bf k}}{\E_{{\bf k}}-\E_{{\bf k-q}}
-\hbar [\omega+i\bar\nu({\bf k},{\bf k-q})]},
$$
then expression (2.4) for electric conductivity and (2.5) for
dielectric perme\-abi\-lity will be transformed to the following form
$$
\sigma_{tr}(\mathbf{q},\omega,\bar\nu)=\dfrac{ie^2N}{m\omega}\big(1+
J_{\bar\nu}+J_\omega \big)
\eqno{(2.6)}
$$
and
$$
\varepsilon_{tr}(\mathbf{q},\omega,\bar\nu)=1-\dfrac{\omega_p^2}{\omega^2}
\big(1+J_{\bar\nu}+J_\omega \big).
\eqno{(2.7)}
$$

Integrals $J_{\bar\nu}$ and  $J_\omega$ we can transform to the following
form
$$
J_{\bar\nu}=\dfrac{i\hbar^2}{8\pi^3mN}
\int\Bigg[\dfrac{\bar\nu({\bf k},{\bf k-q})}{[\omega+i\bar\nu({\bf k},{\bf k-q})]
\{\E_{{\bf k}}-\E_{{\bf k-q}}\}}-
$$
$$
-\dfrac{\bar\nu({\bf k+q},{\bf k})}{[\omega+i\bar\nu({\bf k+q},{\bf k})]
\{\E_{{\bf k+q}}-\E_{{\bf k}}\}}
\Bigg]f_{{\bf k}}{\bf k}_\perp^2d{\bf k}
\eqno{(2.8)}
$$
and
$$
J_{\omega}=\dfrac{\omega\hbar^2}{8\pi^3mN}
\int\Bigg[\dfrac{1}{[\omega+i\bar\nu({\bf k},{\bf k-q})]
\{\E_{{\bf k}}-\E_{{\bf k-q}}-\hbar [\omega+i\bar\nu({\bf k},{\bf k-q})]\}}-
$$
$$
-\dfrac{1}{[\omega+i\bar\nu({\bf k+q},{\bf k})]
\{\E_{{\bf k+q}}-\E_{{\bf k}}-\hbar [\omega+i\bar\nu({\bf k+q},{\bf k})]\}}
\Bigg]f_{{\bf k}}{\bf k}_\perp^2d{\bf k}.
\eqno{(2.9)}
$$ \bigskip

\begin{center}
\bf 3. Non--degenerate plasma
\end{center}

Instead of the vector $\mathbf{k}$ we will enter the dimensionless
vector $\mathbf {K}$ by following equality
$\mathbf{K}=\dfrac{\mathbf{k}}{k_T}$, \quad
$k_T=\dfrac{p_T}{\hbar}$, where $k_T$ is the thermal wave number,
$p_T=mv_T$ is the thermel electron momentum,
$$
v_T=\dfrac{1}{\sqrt{\beta}}=\sqrt{\dfrac{2k_BT}{m}}
$$
is the thermal electron velocity, $k_B$ is the Boltzmann
constant, $T$ is the plasma temperature.

Then
$$
(k^2-k_x^2)d^3k=k_T^5(K^2-K_x^2)d^3K=k_T^5K_\perp^2d^3K,
$$
where
$$
K_\perp^2=K^2-K_x^2=K_y^2+K_z^2.
$$

Further we will consider the case of non-degenerate plasmas.
Then we have
$$
\Big(\dfrac{mv_T}{\hbar}\Big)^3 \equiv \Big(\dfrac{p_T}{\hbar}\Big)^3\equiv
k_T^3=\dfrac{\pi^2}{f_2(\alpha)}N.
$$

Here
$$
f_2(\alpha)=\int\limits_{0}^{\infty}f_F(K,\alpha)K^2dK=
\int\limits_{0}^{\infty}\dfrac{K^2dK}{1+e^{K^2-\alpha}},
$$
where
$$
f_F(K,\alpha)=\dfrac{1}{1+e^{K^2-\alpha}}.
$$

Hence,
$$
(k^2-k_x^2)d^3k=
\dfrac{\pi^2}{f_2(\alpha)}N\dfrac{m^2v_T^2}{\hbar^{2}}(K^2-K_x^2)d^3K.
$$

Energy $\E_{\mathbf {k}} $ we will express through thermal energy. We have

$$
\E_{\mathbf{k}}=\dfrac{\hbar^2\mathbf{k}^2}{2m}=\dfrac{\hbar^2k_T^2}{2m}
\mathbf{K}^2=\dfrac{p_T^2}{2m}\mathbf{K}^2=
\E_T\mathbf{K}^2\equiv \E_{\mathbf{K}}.
$$

Here $\E_T=\dfrac{p_T^2}{2m}$ is the thermal electron energy.

Besides,
$$
k_BT=\dfrac{2k_BT}{m}\cdot\dfrac{m}{2}=\dfrac{mv_T^2}{2}=\E_T,
$$
$$
\dfrac{\E_{{\bf k}}}{k_BT}={\bf K}^2=K^2,\qquad K=|{\bf K}|.
$$
Absolute  Fermi---Dirac's distribution  $f_{\mathbf{k}}$
for non-degenerate plasma is dist\-ri\-bu\-tion
$$
f_{\mathbf{k}}=\dfrac{1}{1+\exp\Big(\dfrac{\E_{{\bf k}}}{k_BT}-\alpha\Big)}=
\dfrac{1}{1+e^{{\bf K}^2-\alpha}}=f_{{\bf K}}\equiv f_{F}(K,\alpha).
$$

Calculation of transversal electric conductivity can be spent on any of
formulas (2.4) -- (2.7).

In the same way we receive
$$
\E_{\mathbf{k-q}}=\dfrac{\hbar^2(k_T\mathbf{K}-\mathbf{q})^2}{2m}=
\dfrac{\hbar^2k_T^2}{2m}\Big(\mathbf{K}-\dfrac{\mathbf{q}}{k_T}\Big)^2.
$$

Further we introduce dimensionless wave vector
$\mathbf{Q}=\dfrac{\mathbf{q}}{k_T}$. Then
$$
\E_{\mathbf{k-q}}=\dfrac{\hbar^2k_T^2}{2m}\Big(\mathbf{K}-\mathbf{Q}\Big)^2=
\E_T(\mathbf{K-Q})^2=\E_{{\bf K-Q}}.
$$

We notice that
$$
\E_{\mathbf{K}}-\E_{\mathbf{K-Q}}=\E_T\mathbf{K}^2-\E_F(\mathbf{K-Q})^2=
\E_F[2K_xQ-Q^2]=
$$
$$
=2Q\E_T(K_x-\dfrac{Q}{2}).
$$

Besides
$$
\E_{\mathbf{K}}-\E_{\mathbf{K-Q}}-\hbar [\omega+i\bar\nu({\bf K},{\bf K-Q})]=
2\E_TQ\Big(K_x-\dfrac{z^-}{Q}-\dfrac{Q}{2}),
$$
$$
\E_{\mathbf{K+Q}}-\E_{\mathbf{K}}-\hbar [\omega+i\bar\nu({\bf K+Q},{\bf K})]=
2\E_TQ\Big(K_x-\dfrac{z^+}{Q}+\dfrac{Q}{2}),
$$
where
$$
z^{\pm}=x+iy^{\pm},\qquad x=\dfrac{\omega}{k_Tv_T},
$$
$$
y^-=\dfrac{\bar\nu({\bf K},{\bf K-Q})}{k_Tv_T},\qquad
y^+=\dfrac{\bar\nu({\bf K+Q},{\bf K})}{k_Tv_T}.
$$

Now formulas (2.4) and (2.5) we can rewrite in the form
$$
\sigma_{tr}(Q,x,y)=\dfrac{ie^2N}{m\omega}\Big(1+J_\omega+J_{\bar\nu}\Big)
\eqno{(3.1)}
$$
and
$$
\varepsilon_{tr}(Q,x,y)=1-\dfrac{x_p^2}{x^2}\Big(1+J_\omega+J_{\bar\nu}\Big),
\qquad x_p=\dfrac{\omega_p}{k_Tv_T}.
\eqno{(3.2)}
$$

In formulas (3.1) and (3.2) there are the following designations
$$
J_\omega=J_\omega(Q,x,y)=\dfrac{1}{8\pi f_2(\alpha)Q}\int\Bigg[\dfrac{x}{(x+iy^-)
(K_x-z^-/Q-Q/2)}-$$$$-
\dfrac{x}{(x+iy^+)(K_x-z^+/Q+Q/2)}\Bigg]f_{{\bf K}}{\bf K}_\perp^2d^3K,
$$
$$
J_{\bar \nu}(Q,x,y)=\dfrac{1}{8\pi f_2(\alpha)Q}
\int\Bigg[\dfrac{iy^-}{(x+iy^-)
(K_x-Q/2)}-$$$$-\dfrac{iy^+}{(x+iy^+)(K_x+Q/2)}\Bigg]f_{{\bf K}}{\bf
K}_\perp^2d^3K.
$$

\begin{center}
\bf 4. Frequency of collisions is proportional to the module
of a wave vector
\end{center}

Let's consider the special case, when frequency of collisions
is proportional to the module of a wave vector
$$
\nu({\bf k})=\nu_0|{\bf k}|.
$$

Then
$$
\bar\nu({\bf k,k-q})=\dfrac{\nu({\bf k})+\nu({\bf k-q})}{2}=
\dfrac{\nu_0}{2}\Big(|{\bf k}|+|{\bf k-q}|\Big),
$$
and
$$
\bar\nu({\bf k+q,k})=\dfrac{\nu({\bf k+q})+\nu({\bf k})}{2}=
\dfrac{\nu_0}{2}\Big(|{\bf k+q}|+|{\bf k}|\Big).
$$

The quantity $ \nu_0$ we take in the form $\nu_0=\dfrac{\nu}{k_T}$, where
$k_T$ is the thermal wave number, $k_T=\dfrac{mv_T}{\hbar}$,
$ \hbar $ is the Planck's constant, $v_T$ is the thermal electron velocity. Now
we have
$$
\nu({\bf k})=\dfrac{\nu}{k_T}|{\bf k}|.
\eqno{(4.1)}
$$

Let's notice, that  at $k=k_T$:
$ \nu (k_T) = \nu $. So, further in previous formulas frequency
collisions according to (4.1) it is equal
$$
\bar\nu({\bf k,k-q})=\dfrac{\nu}{2k_T}
\big(|{\bf k}|+|{\bf k-q}|\big)=\dfrac{\nu}{2}\Big(|{\bf K}|+|{\bf K-Q}|\Big)=
\bar\nu({\bf K,K-Q}),
$$
$$
\bar\nu({\bf k+q,k})=\dfrac{\nu}{2k_T}
\big(|{\bf k+q}|+|{\bf k}|\big)=\dfrac{\nu}{2}\Big(|{\bf K+Q}|+|{\bf K}|\Big)=
\bar\nu({\bf K+Q,K}).
$$

Hence, quantities $z^{\pm}$ are equal
$$
z^{\pm}=x+iy\rho^{\pm}, \qquad y=\dfrac{\nu}{k_Fv_F},
$$
$$
\rho^{\pm}=\dfrac{1}{2}\Big(\sqrt{K_x^2+K_y^2+K_z^2}+
\sqrt{(K_x\pm Q)^2+K_y^2+K_z^2}\Big).
$$

Now integrals $J_{\bar\nu} $ and $J_\omega $ are accordingly equal
$$
J_{\bar\nu}=\dfrac{iy}{8\pi Qf_2(\alpha)}\int\Bigg(\dfrac{\rho^-}{(x+iy\rho^-)
(K_x-Q/2)}-\hspace{5cm}$$$$\hspace{3cm}-\dfrac{\rho^+}{(x+iy\rho^+)
(K_x+Q/2)}\Bigg)f_{{\bf K}}{\bf K}_\perp^2d^3K,
\eqno{(4.2)}
$$
and
$$
J_{\omega}=\dfrac{x}{8\pi Qf_2(\alpha)}\int\Bigg(\dfrac{1}{(x+iy\rho^-)
(K_x-z^-/Q-Q/2)}-\hspace{4cm}$$$$\hspace{4cm}-\dfrac{1}{(x+iy\rho^+)
(K_x-z^+/Q+Q/2)}\Bigg)f_{{\bf K}}{\bf K}_\perp^2d^3K.
\eqno{(4.3)}
$$

Three-dimensional integrals (4.2) and (4.3) after passing to polar
coor\-di\-na\-tes in a plane $(K_y, K_z) $ are easily reduced to the double
$$
J_{\bar\nu}=\dfrac{iy}{4Qf_2(\alpha)}\int\limits_{-\infty}^{+\infty}dK_x
\int\limits_{0}^{\infty}\Bigg(\dfrac{\rho^-}{(x+iy\rho^-)
(K_x-Q/2)}-\hspace{5cm}
$$
$$
\hspace{3cm}-\dfrac{\rho^+}{(x+iy\rho^+)
(K_x+Q/2)}\Bigg)f_F(K_x,r,\alpha)r^3dr,
\eqno{(4.4)}
$$
and
$$
J_{\omega}=\dfrac{x}{4 Qf_2(\alpha)}\int\limits_{-\infty}^{+\infty}dK_x
\int\limits_{0}^{\infty}\Bigg(\dfrac{1}{(x+iy\rho^-)
(K_x-z^-/Q-Q/2)}-\hspace{4cm}
$$
$$
\hspace{4cm}-\dfrac{1}{(x+iy\rho^+)
(K_x-z^+/Q+Q/2)}\Bigg)f_F(K_x,r,\alpha)r^3dr.
\eqno{(4.5)}
$$
Here
$$
f_F(K_x,r,\alpha)=\dfrac{1}{1+e^{K_x^2+r^2-\alpha}}=
f_F(K,\alpha),\qquad r^2=K_y^2+K_z^2,
$$
$$
\rho^{\pm}=\rho^{\pm}(K_x,r)=\dfrac{1}{2}\Big(\sqrt{K_x^2+r^2}+
\sqrt{(K_x\pm Q)^2+r^2}\Big).
$$

Let's notice, that in case of constant frequency of collisions
$\rho^{\pm} =1$ and formulas (4.4) and (4.5) pass in the following
$$
J_\omega=\dfrac{x}{4(x+iy)f_2(\alpha)}I(Q,z),
$$
where
$$
I(Q,z)=\int\limits_{-\infty}^{\infty}\dfrac{f_3(K_x,\alpha)dK_x}
{(K_x-z/Q)^2-(Q/2)^2},
$$
$$
f_3(K_x,\alpha)=\int\limits_{0}^{\infty}\ln(1+e^{K_x^2+r^2-\alpha})rdr,
$$
$$
J_{\nu}=\dfrac{iy}{4(x+iy)}I(Q,0), \qquad z=x+iy.
$$

By means of these expressions we receive known formulas for
electric conductivity and dielectric permeability
of quantum collisional degenerate plasmas with constant
frequency of collisions of particles \cite{Trans2}
$$
\dfrac{\sigma_{tr}}{\sigma_0}=\dfrac{iy}{x}\Big[1+\dfrac{1}{4f_2(\alpha)}
\dfrac{xI(Q,z)+iyI(Q,0)}{x+iy}\Big]
\eqno{(4.6)}
$$
and
$$
\varepsilon_{tr}=1-\dfrac{\omega_p^2}{\omega^2}\Big[1+\dfrac{1}{4f_2(\alpha)}
\dfrac{xI(Q,z)+iyI(Q,0)}{x+iy}\Big].
\eqno{(4.7)}
$$

\begin{center}
\bf  5. Maxwellian plasma
\end{center}

Dimensionless vectors ${\bf K}$ and ${\bf Q}$ are entered as well as in the
case of non-degenerate plasmas, however now
$$
f_{{\bf k}}=4\pi^{3/2}\dfrac{N}{k_T^3}
\exp\Big(-\dfrac{\E_{{\bf k}}}{\E_T}\Big)=
4\pi^{3/2}\dfrac{N}{k_T^3}e^{-{\bf K}^2}\equiv f_{{\bf K}},
$$
thus
$$
N \equiv f_{{\bf k}}\dfrac{2d^3{\bf p}}{(2\pi \hbar)^3},\qquad
{\bf k}_\perp^2\,d{\bf k}=k_T^5{\bf K}_\perp^2\,d^3K.
$$

Electric conductivity and
dielectric permeability is calculated again under formulas (3.1) and (3.2)
$$
\sigma_{tr}(Q,x,y)=\dfrac{ie^2N}{m\omega}\Big(1+J_\omega+J_{\bar\nu}\Big)
$$
and
$$
\varepsilon_{tr}(Q,x,y)=1-\dfrac{x_p^2}{x^2}\Big(1+J_\omega+J_{\bar\nu}\Big)
$$

In these formulas
$$
J_\omega=\dfrac{x}{2\pi^{3/2}Q}
\int\Bigg(\dfrac{1}{(x+iy\rho^-)
(K_x-z^-/Q-Q/2)}-\hspace{4cm}$$$$\hspace{4cm}-\dfrac{1}{(x+iy\rho^+)
(K_x-z^+/Q+Q/2)}\Bigg)f_{{\bf K}}{\bf K}_\perp^2d^3K,
$$
$$
J_{\bar\nu}=\dfrac{iy}{2\pi^{3/2}Q}\int\Bigg(\dfrac{\rho^-}{(x+iy\rho^-)
(K_x-Q/2)}-\hspace{5cm}$$$$\hspace{3cm}-\dfrac{\rho^+}{(x+iy\rho^+)
(K_x+Q/2)}\Bigg)f_{{\bf K}}{\bf K}_\perp^2d^3K.
$$

After integration in the plane $(K_y,K_z)$ previous formulas
have the following form
$$
J_\omega=\dfrac{x}{\sqrt{\pi}Q}
\int\limits_{-\infty}^{+\infty}e^{-K_x^2}dK_x\int\limits_{0}^{\infty}\Bigg(\dfrac{1}{(x+iy\rho^-)
(K_x-z^-/Q-Q/2)}-\hspace{4cm}$$$$\hspace{4cm}-\dfrac{1}{(x+iy\rho^+)
(K_x-z^+/Q+Q/2)}\Bigg)e^{-r^2}r^3dr,
$$

$$
J_{\bar\nu}=\dfrac{iy}{\sqrt{\pi}Q}
\int\limits_{-\infty}^{+\infty}e^{-K_x^2}dK_x\int\limits_{0}^{\infty}
\Bigg(\dfrac{\rho^-}{(x+iy\rho^-)
(K_x-Q/2)}-\hspace{5cm}$$$$\hspace{3cm}-\dfrac{\rho^+}{(x+iy\rho^+)
(K_x+Q/2)}\Bigg)e^{-r^2}r^3dr.
$$

In case of constant frequency of collisions we have
$\rho^{\pm}=1$, thus
$$
J_\omega=\dfrac{x}{x+iy}I(Q,z),\qquad
I(Q,z)=\dfrac{1}{2\sqrt{\pi}}\int\limits_{-\infty}^{\infty}
\dfrac{e^{-K_x^2}\,dK_x}{(K_x-z/Q)^2-(Q/2)^2},
$$
$$
J_\nu=\dfrac{iy}{x+iy}I(Q,0), \qquad
I(Q,0)=\dfrac{1}{2\sqrt{\pi}}\int\limits_{-\infty}^{\infty}
\dfrac{e^{-K_x^2}\,dK_x}{K_x^2-(Q/2)^2}.
$$

To these formulas it is possible to come and other method.
Really, in formulas (3.6) and (3.7) we will pass to the limit at $\alpha\to
-\infty$. For this purpose we will calculate the following limit
$$
\lim\limits_{\alpha\to -\infty}\dfrac{f_3(K_x,\alpha)}{f_2(\alpha)}=
\dfrac{2}{\sqrt{\pi}}e^{-K_x^2}.
\eqno{(5.1)}
$$

By means of equality (5.1) from previous formulas (3.6) and (3.7)
we receive known formulas for quantum maxwellian plasmas with constant
frequency of collisions
$$
\sigma_{tr}(x,y,Q)=i\sigma_0\dfrac{y}{x}\Big[1+\dfrac{xI(Q,z)+iyI(Q,0)}
{x+iy}\Big]
$$
and
$$
\varepsilon_{tr}(x,y,Q)=1-\dfrac{x_p^2}{x^2}\Big[1+\dfrac{xI(Q,z)+iyI(Q,0)}
{x+iy}\Big].
$$

Here
$$
I(Q,z)=\dfrac{1}{2\sqrt{\pi}}\int\limits_{-\infty}^{\infty}
\dfrac{e^{-K_x^2}\,dK_x}{(K_x-z/Q)^2-(Q/2)^2}.
$$

On Figs. 1 -- 12 we will present comparison real and imaginary
parts of dielectric function, $x_p=1$. Figures 1-6 are devoted
to maxwellian plasma, and figures 7-12 are devoted to non-degenerate
plasma.

\begin{center}
\bf 5. Conclusion
\end{center}

In the present work formulas for the transversal electric
conductivity and dielectric permeability into quantum non-degenerate
and maxwellian collisional
plasma are deduced.
Frequency of collisions of particles depends arbitrarily on a wave vector.
The special case of frequency of collisions
proportional to the module of a wave vector is considered.
The graphic analysis of the real and imaginary parts  of
dielectric function is made.

\begin{figure}[h]\center
\includegraphics[width=15.0cm, height=10cm]{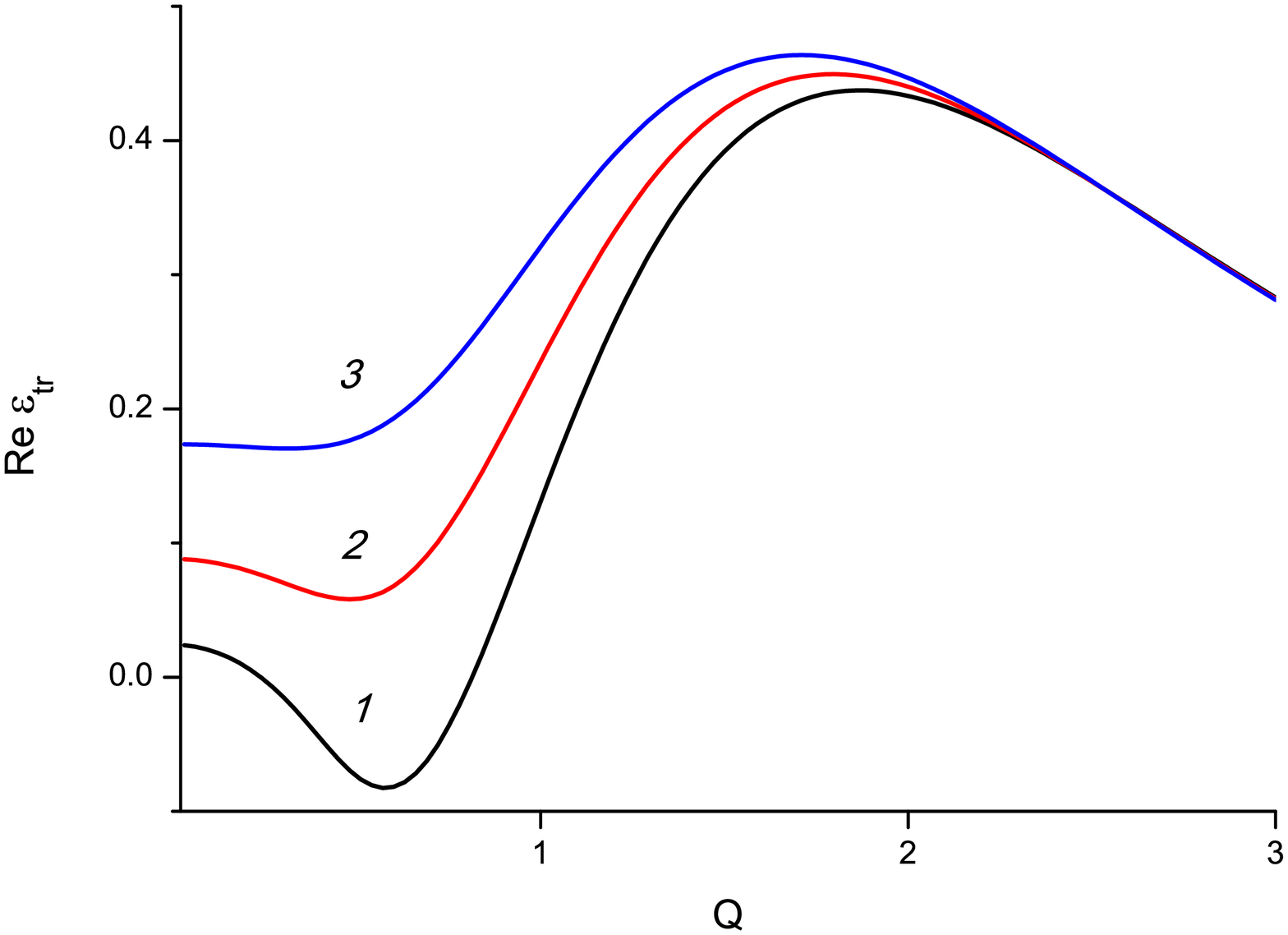}
\center{Fig. 1. Real part of dielectric function,
$x=1$. Curves 1,2,3 correspond to values $y=0.1,0.2,0.3$.}
\includegraphics[width=15.0cm, height=10cm]{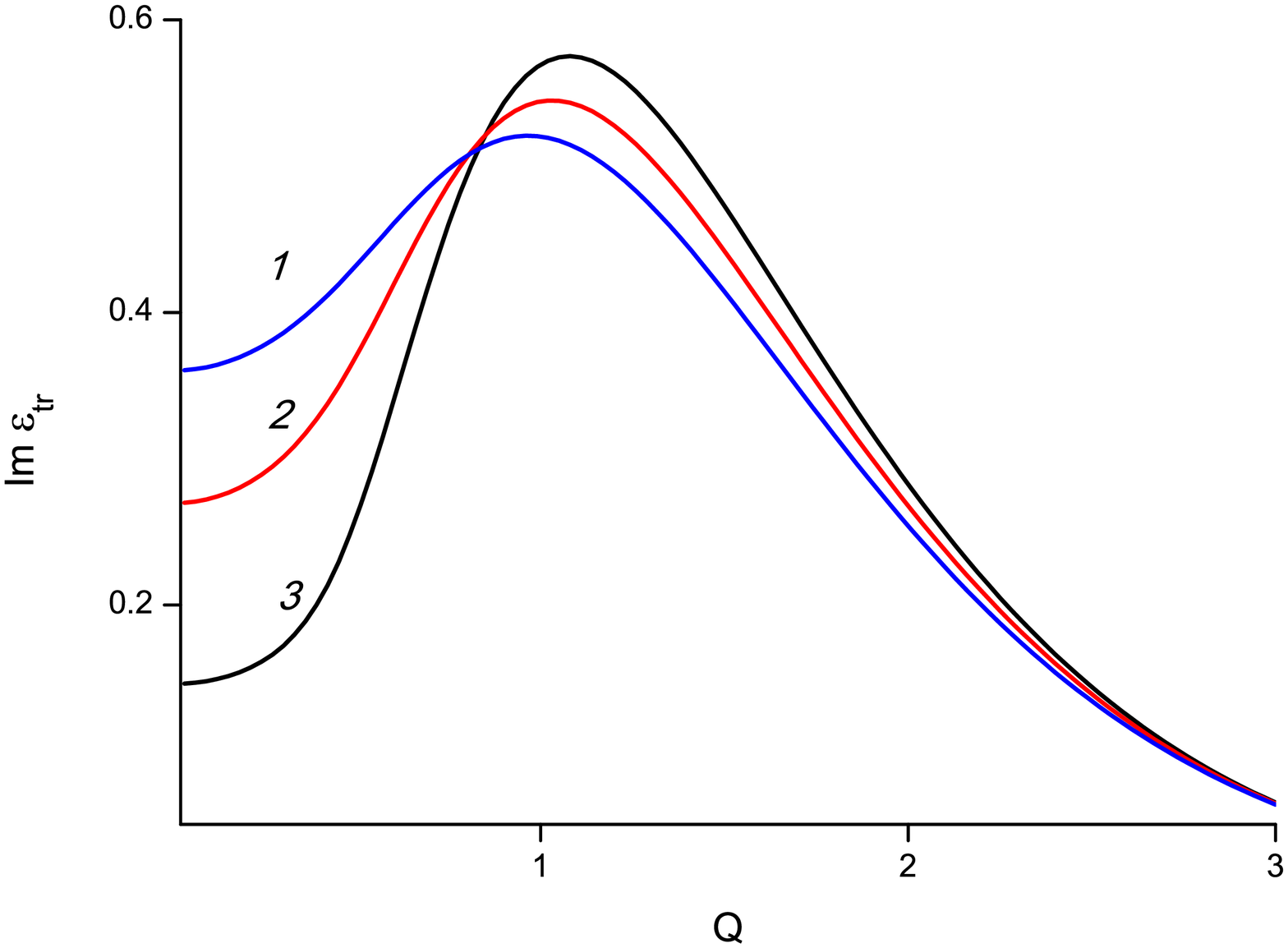}
\center{Fig. 2. Imaginare part of dielectric function,
$x=1$. Curves 1,2,3 correspond to values $y=0.1,0.2,0.3$.}
\end{figure}
\begin{figure}[h]\center
\includegraphics[width=15.0cm, height=10cm]{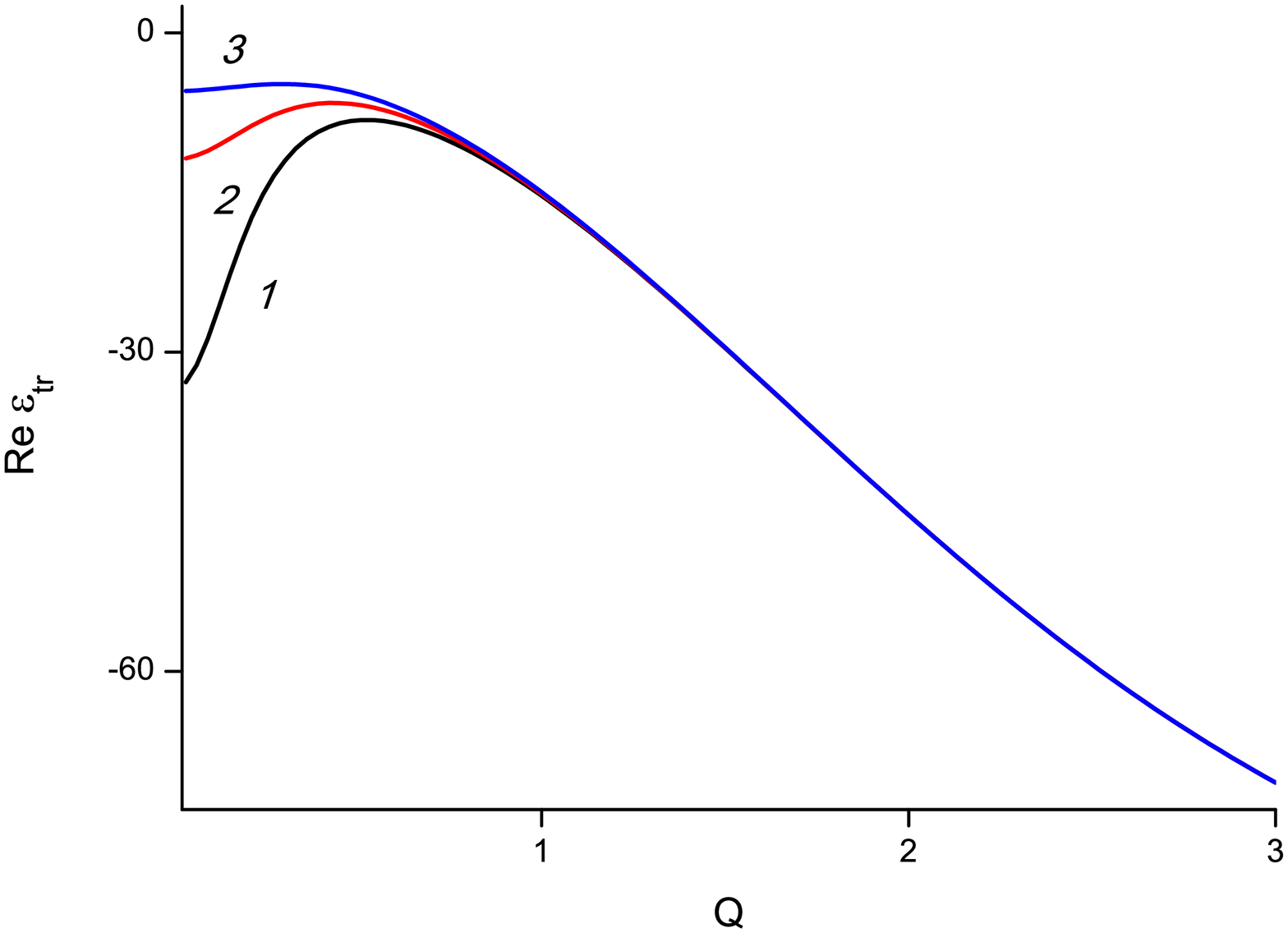}
\center{Fig. 3. Real part of dielectric function,
$x=0.1$. Curves 1,2,3 correspond to values $y=0.1,0.2,0.3$.}
\includegraphics[width=15.0cm, height=10cm]{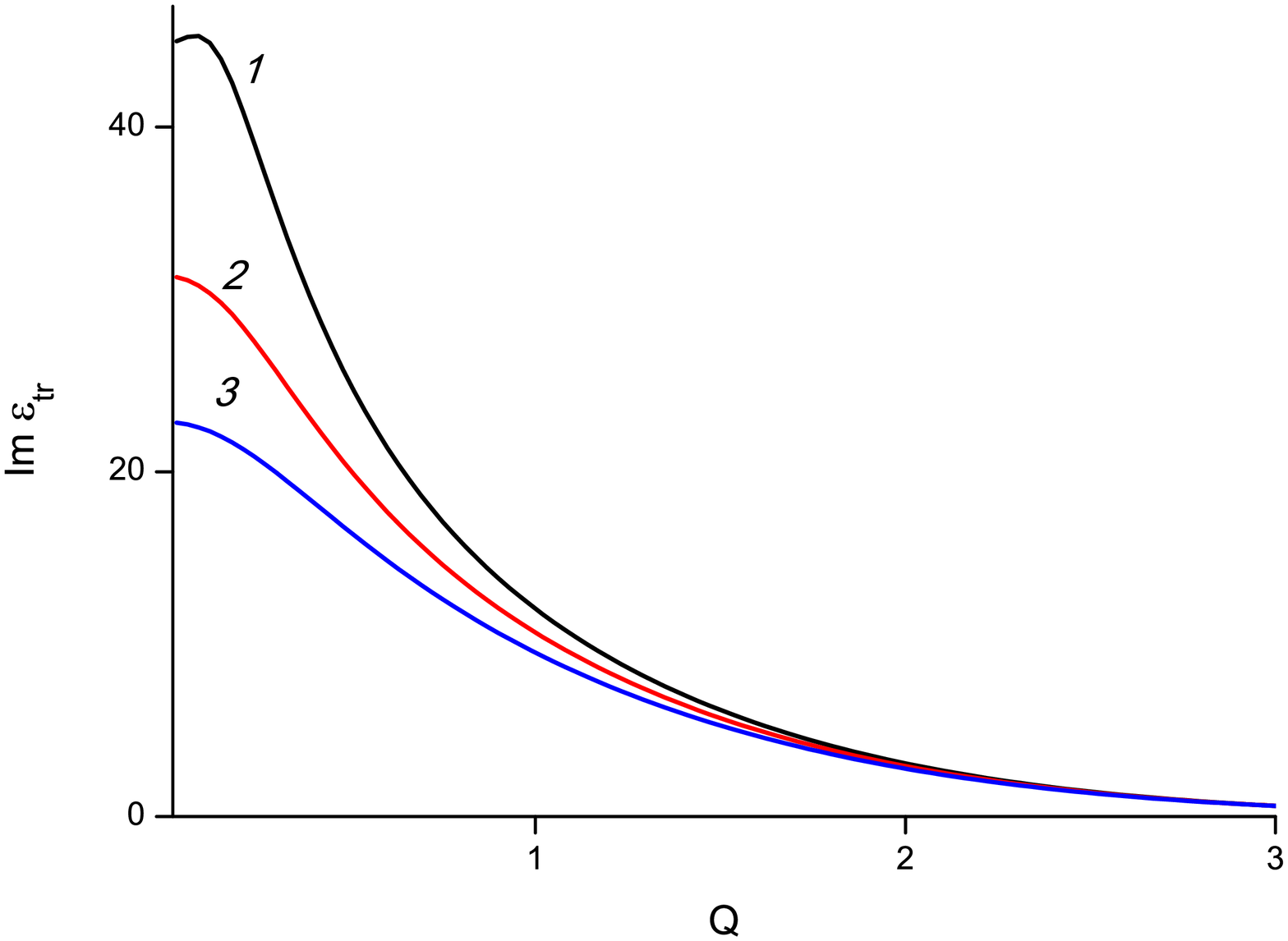}
\center{Fig. 4. Imaginare part of dielectric function,
$x=0.1$. Curves 1,2,3 correspond to values $y=0.1,0.2,0.3$.}
\end{figure}
\begin{figure}[h]\center
\includegraphics[width=15.0cm, height=10cm]{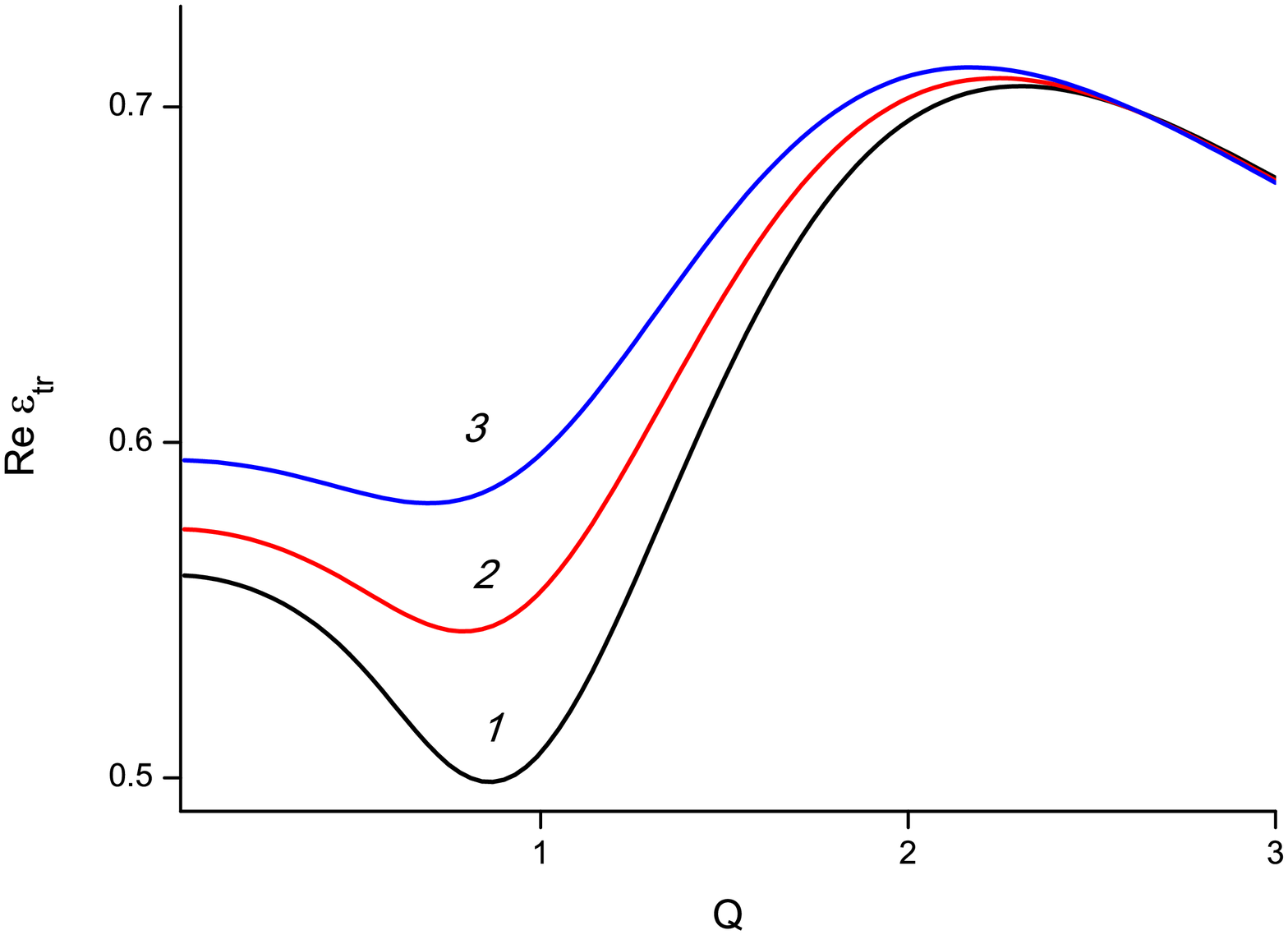}
\center{Fig. 5. Real part of dielectric function,
$x=1.5$. Curves 1,2,3 correspond to values $y=0.1,0.2,0.3$.}
\includegraphics[width=15.0cm, height=10cm]{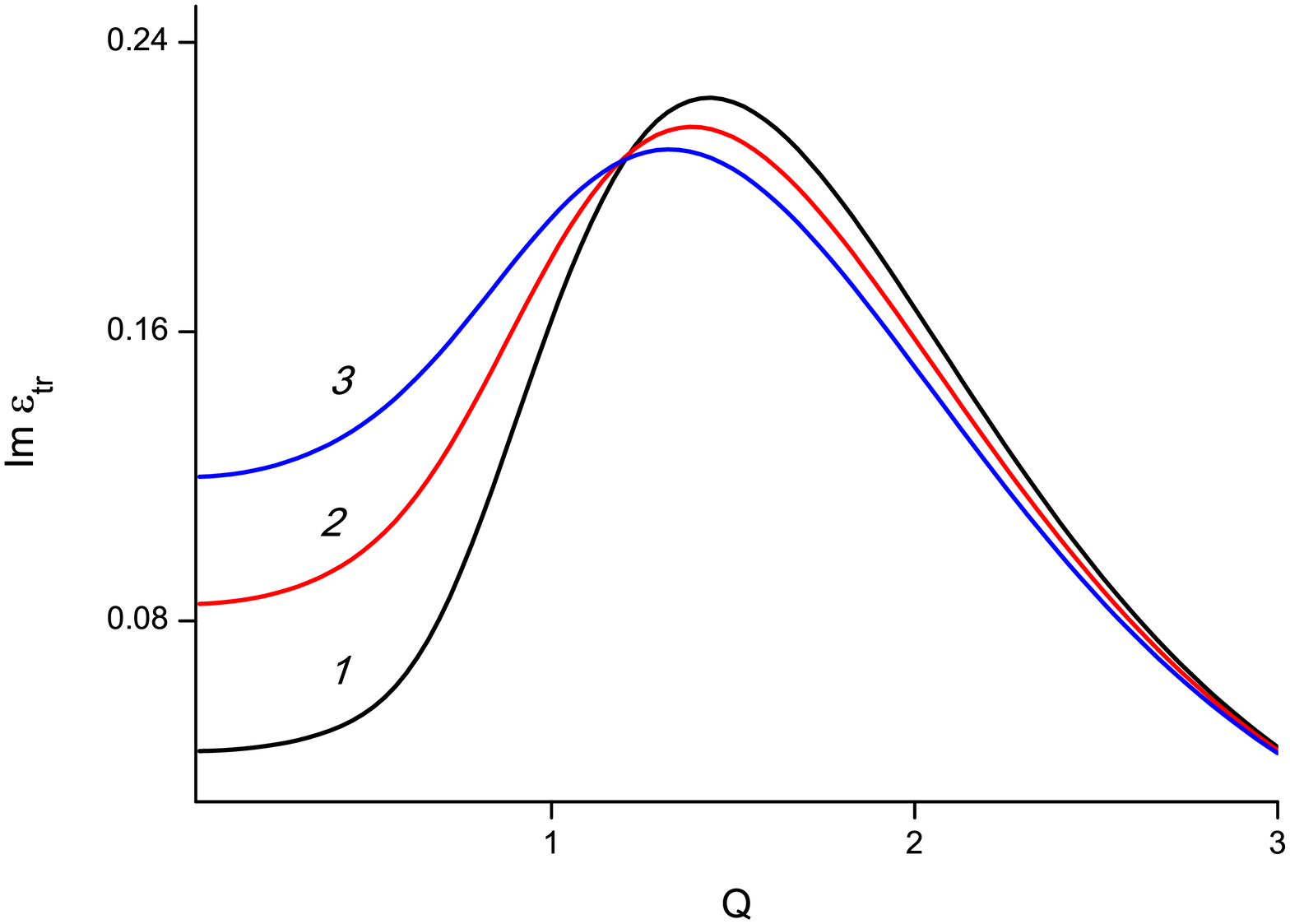}
\center{Fig. 6. Imaginare part of dielectric function,
$x=1.5$. Curves 1,2,3 correspond to values $y=0.1,0.2,0.3$.}
\end{figure}

\begin{figure}[h]\center
\includegraphics[width=15.0cm, height=10cm]{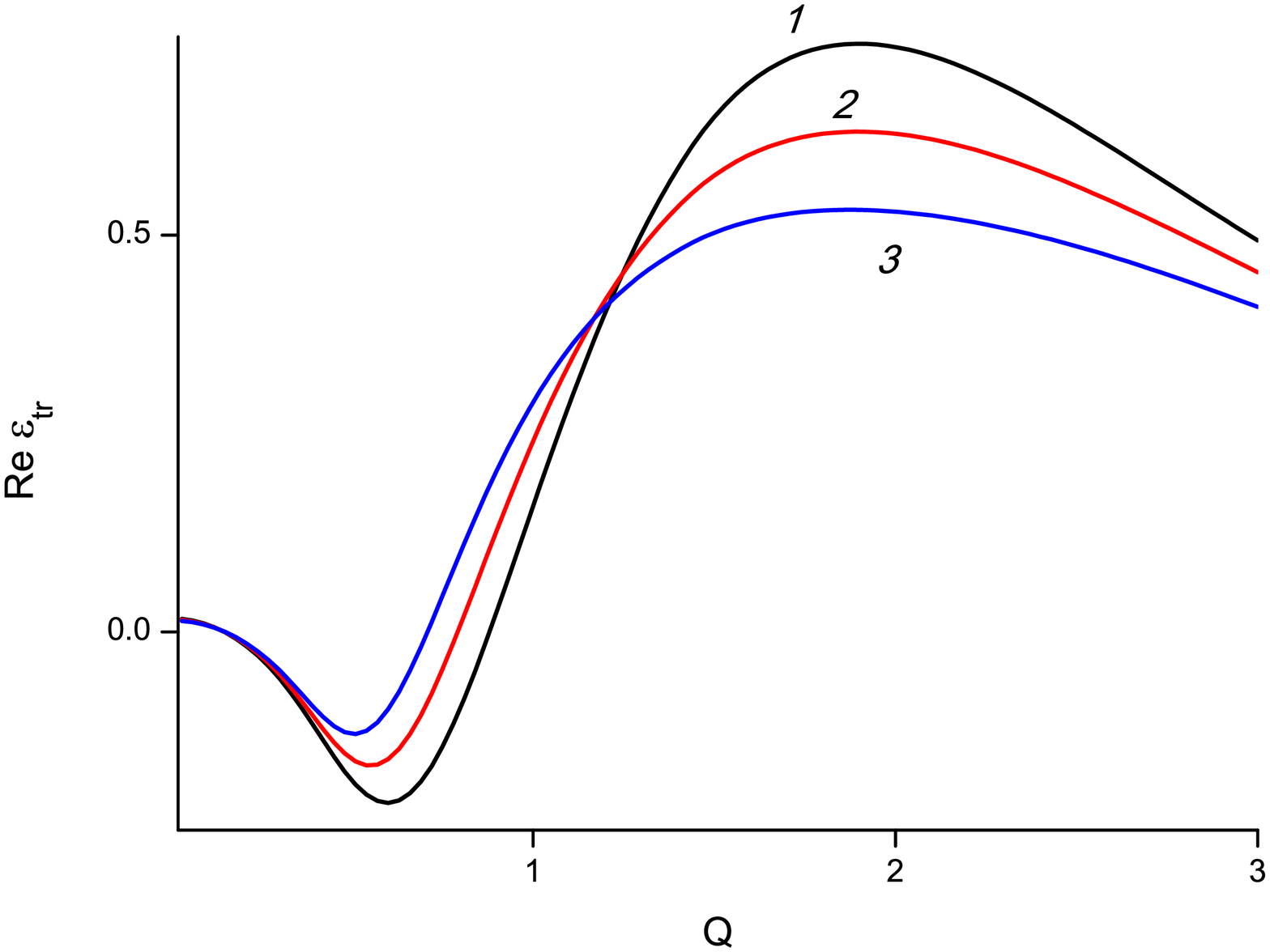}
\center{Fig. 7. Real part of dielectric function,
$x=1,y=0.1$. Curves 1,2,3 correspond to values $\alpha=-2,0,+1$.}
\includegraphics[width=15.0cm, height=10cm]{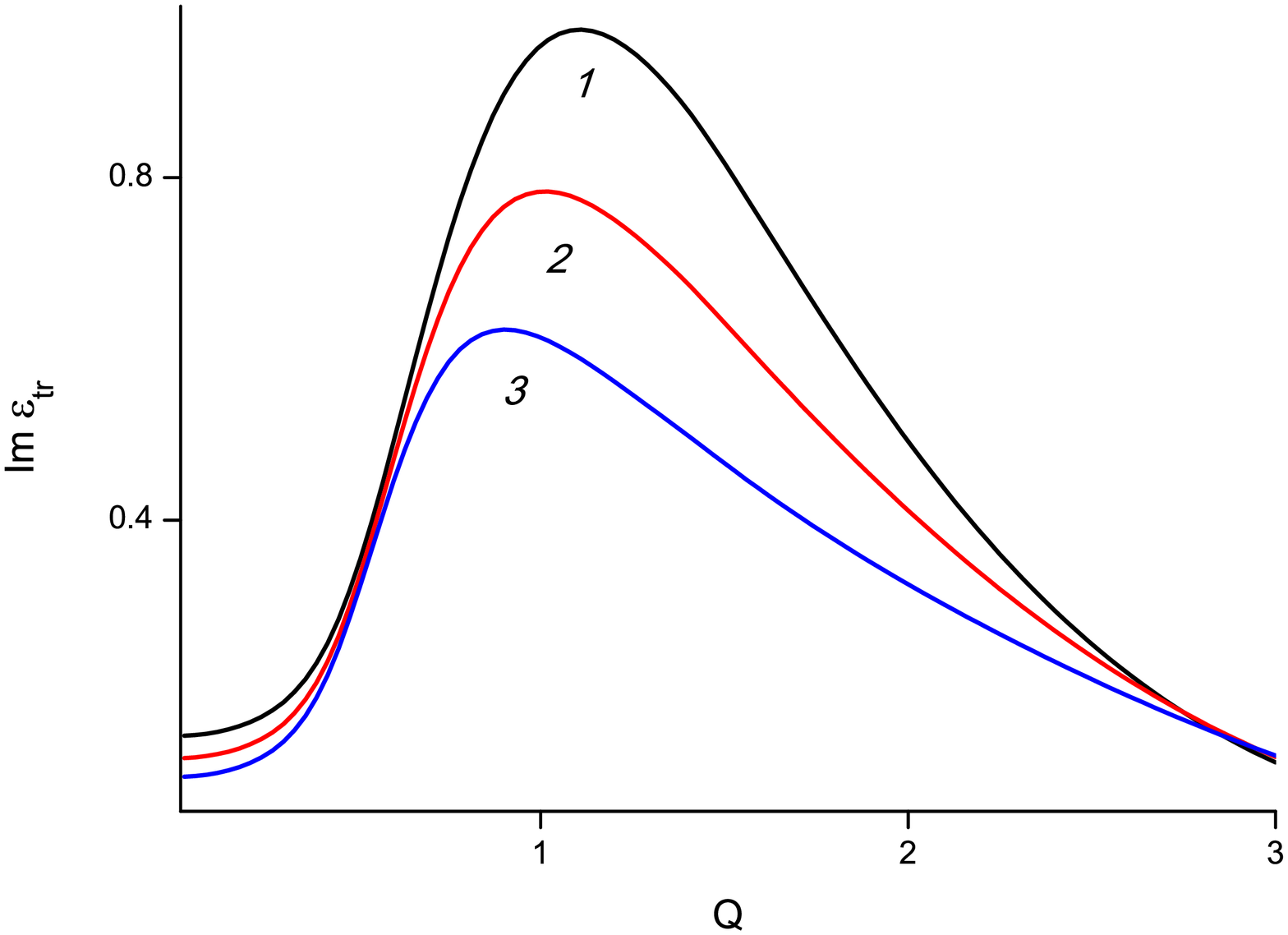}
\center{Fig. 8. Imaginare part of dielectric function,
$x=1,y=0.1$. Curves 1,2,3 correspond to values $\alpha=-2,0,+1$.}
\end{figure}
\begin{figure}[h]\center
\includegraphics[width=15.0cm, height=10cm]{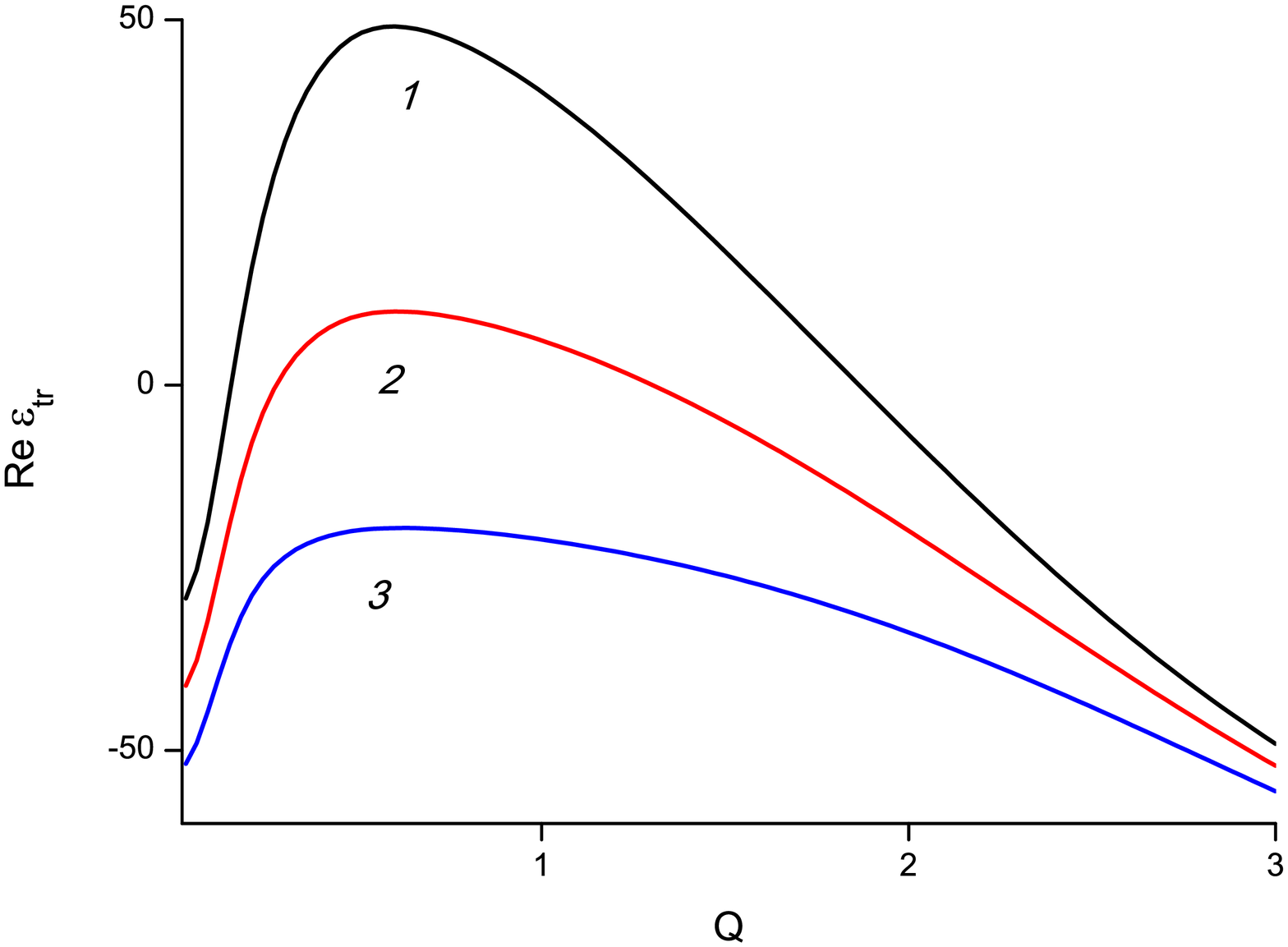}
\center{Fig. 9. Real part of dielectric function,
$x=0.1,y=0.1$. Curves 1,2,3 correspond to values $\alpha=-2,0,+1$.}
\includegraphics[width=15.0cm, height=10cm]{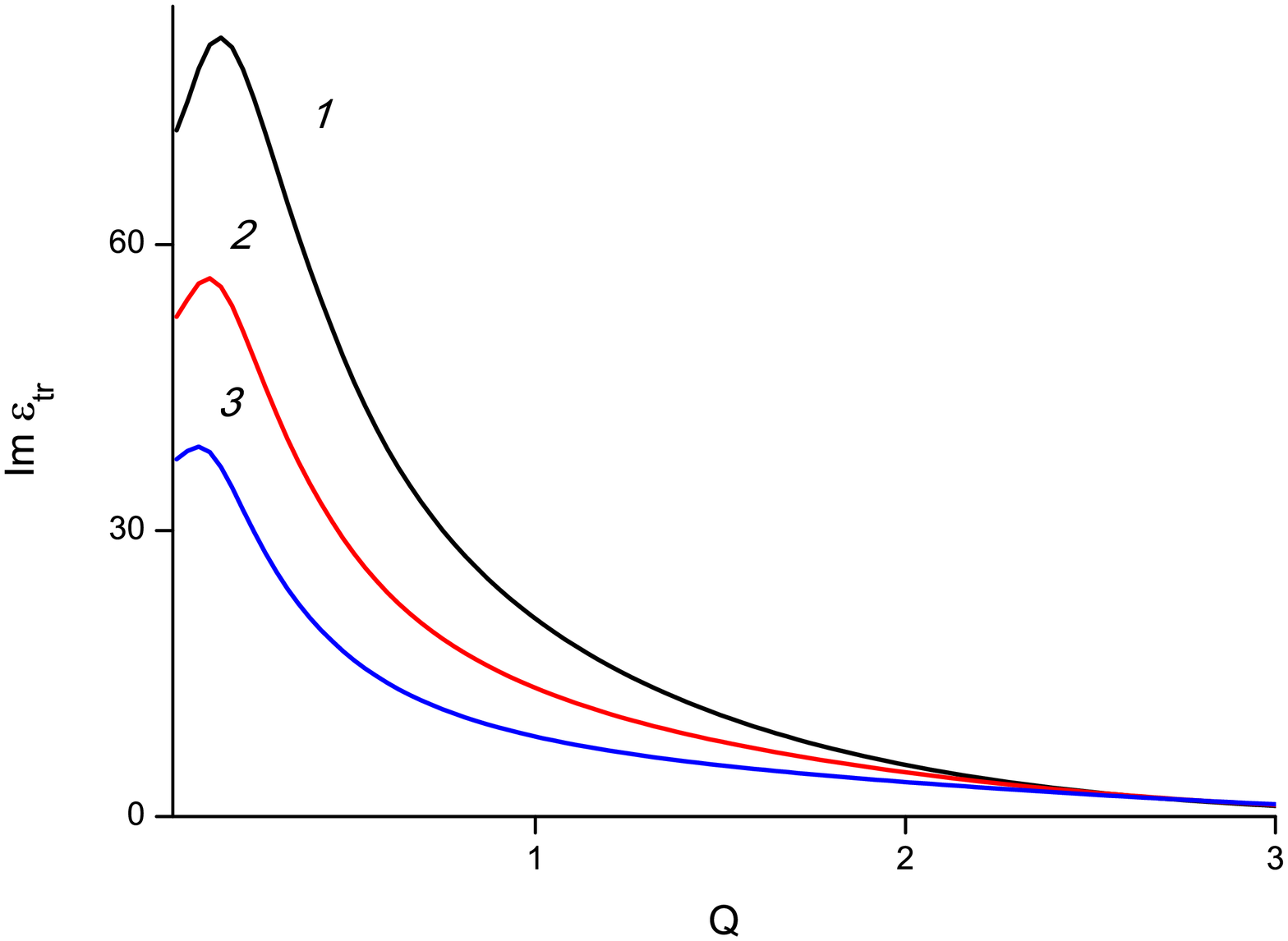}
\center{Fig. 10. Imaginare part of dielectric function,
$x=0.1$. Curves 1,2,3 correspond to values $\alpha=-2,0,+1$.}
\end{figure}
\begin{figure}[h]\center
\includegraphics[width=15.0cm, height=10cm]{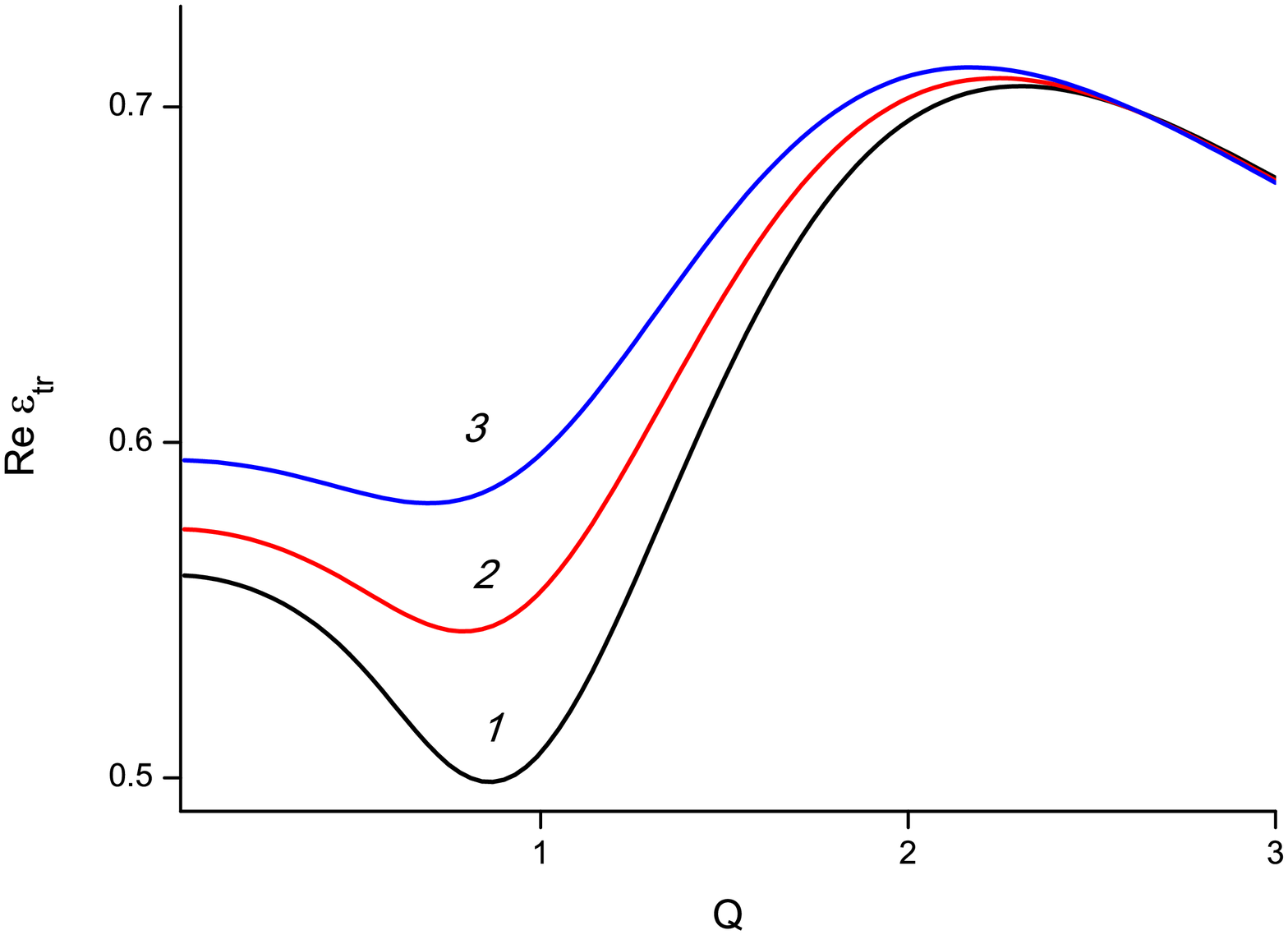}
\center{Fig. 11. Real part of dielectric function,
$x=1.5,y=0.1$. Curves 1,2,3 correspond to values $\alpha=-2,0,+1$.}
\includegraphics[width=15.0cm, height=10cm]{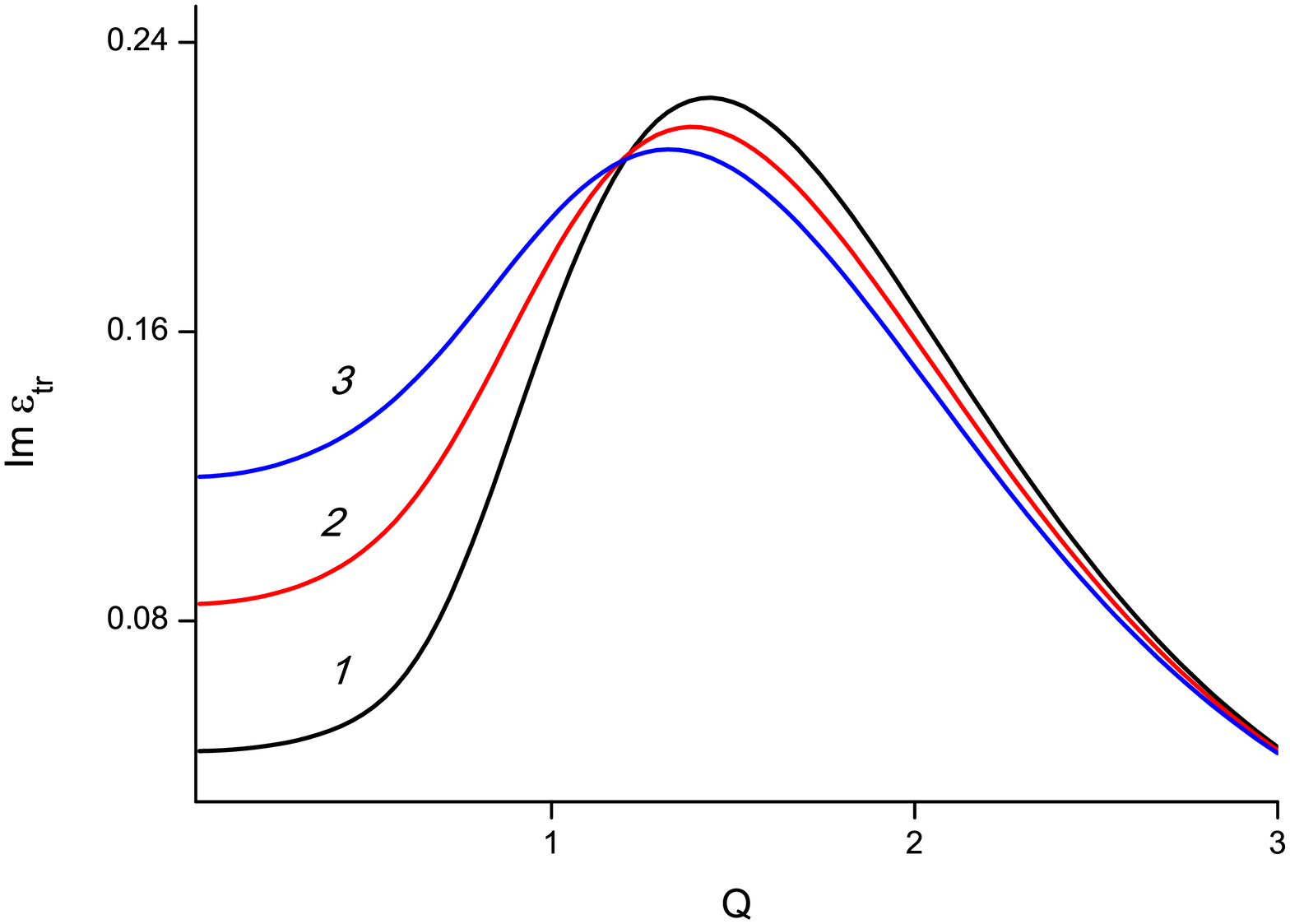}
\center{Fig. 12. Imaginare part of dielectric function,
$x=1.5,y=0.1$. Curves 1,2,3 correspond to values $\alpha=-2,0,+1$.}
\end{figure}

\end{document}